Conference Paper

# Plagiarism or Productivity? Students Moral Disengagement and Behavioral Intentions to Use ChatGPT in Academic Writing

John Paul P. Miranda[1*], Rhiziel P. Manalese[1], Mark Anthony A. Castro[1], Renen Paul M. Viado[2], Vernon Grace M. Maniago[1], Rudante M. Galapon[2], Jovita G. Rivera[1], Amado B. Martinez, Jr.[1]

1. **Pampanga State University**, Pampanga, Philippines
2. **National University**, Dasmariñas, Philippines

**\* Correspondence:**
John Paul P. Miranda, Pampanga State University, jppmiranda@pampangastateu.edu.ph

## ABSTRACT

This study examined how moral disengagement influences Filipino college students' intention to use ChatGPT in academic writing. The model tested five mechanisms: moral justification, euphemistic labeling, displacement of responsibility, minimizing consequences, and attribution of blame. These mechanisms were analyzed as predictors of attitudes, subjective norms, and perceived behavioral control, which then predicted behavioral intention. A total of 418 students with ChatGPT experience participated. The results showed that several moral disengagement mechanisms influenced students' attitudes and sense of control. Among the predictors, attribution of blame had the strongest influence, while attitudes had the highest impact on behavioral intention. The model explained more than half of the variation in intention. These results suggest that students often rely on institutional gaps and peer behavior to justify AI use. Many believe it is acceptable to use ChatGPT for learning or when rules are unclear. This shows a need for clear academic integrity policies, ethical guidance, and classroom support. The study also recognizes that intention-based models may not fully explain student behavior. Emotional factors, peer influence, and convenience can also affect decisions. The results provide useful insights for schools that aim to support responsible and informed AI use in higher education.

# INTRODUCTION

Several studies have reported that students increasingly use ChatGPT to draft or revise academic assignments (Baek et al., 2024; Stojanov et al., 2024). Filipino students also adopt generative artificial intelligence (Gen AI) to generate ideas, organize essays, and improve written outputs (Espartinez, 2025; Giray & and Aquino, 2024). For example, (Liu et al., n.d.) found that instruction supported by large language models improved writing performance, self-regulated learning strategies, and motivation among EFL students. Despite these benefits, scholars have expressed concerns about how widespread AI use challenges traditional definitions of authorship, originality, and academic integrity (Yeo, 2023), (Chan, 2025). In response, Philippine universities continue to revise and update honor codes and plagiarism policies to address the evolving ethical environment (De La Salle University, 2025; Orbe & Santos, 2022; University of the Philippines, 2023).

Moral disengagement theory provides a framework for understanding how students justify behaviors that may violate academic integrity (Raney, 2020). The theory proposes that individuals rely on cognitive mechanisms such as moral justification, euphemistic labeling, displacement of responsibility, minimizing consequences, and attribution of blame to reduce internal moral conflict when engaging in questionable actions (Raney, 2020; Waqas et al., 2025). Furthermore, (Qu & Wang, 2025) observed that students may interpret AI-generated text as a neutral learning tool, which allows them to avoid feelings of guilt. These cognitive strategies may lead students to consider AI-assisted writing acceptable even when institutional policies emphasize academic honesty.

Although moral disengagement offers valuable insights, understanding students' intentions to use ChatGPT also requires attention to other psychological factors. The Theory of Planned Behavior (TPB) explains that behavioral intention results from attitudes toward the behavior, perceived social norms, and perceived behavioral control. Recent studies have confirmed the relevance of TPB in predicting technology adoption within educational contexts (Chu & Chen, 2016; Naseri & Abdullah, 2024). However, these models have not fully considered how moral reasoning may influence the TPB components. In the Philippine context, (Martinez, 2025) reported rapid AI adoption among students (Villarino, 2025), while (Bozkurt et al., 2024) identified ongoing uncertainty regarding ethical policies for AI use. Despite the growing integration of AI tools in higher education, limited research has explored how moral disengagement mechanisms interact with TPB constructs to shape students' intentions to adopt generative AI tools.

The present study addresses this gap by integrating moral disengagement theory and the Theory of Planned Behavior to examine how Filipino college students form intentions to use ChatGPT for academic writing. The study focuses on students who have experience using ChatGPT in assignments, theses, or feasibility studies. It applies a cross-sectional design and structural equation modeling to estimate the relationships among moral disengagement mechanisms, TPB components, and behavioral intention. The findings aim to clarify the psychosocial processes behind AI adoption and to provide empirical evidence that can inform institutional policies, instructional practices, and ethical guidelines in Philippine higher education.

# THEORETICAL FRAMEWORK AND HYPOTHESES

The present study integrates moral disengagement theory and the Theory of Planned Behavior to examine students' intentions to use ChatGPT for academic writing. Moral disengagement theory explains how individuals rationalize potentially unethical behaviors through cognitive mechanisms (Moore, 2015; Raney, 2020; Zhang et al., 2025) such as moral justification (reframing ChatGPT use as educationally beneficial), euphemistic labeling (softening the ethical implications), displacement of responsibility (shifting blame to external factors like instructors or unclear policies), minimizing consequences (downplaying harm), and attribution of blame (assigning fault to institutional or systemic shortcomings). These mechanisms allow students to cognitively justify their ChatGPT use despite academic integrity guidelines. The TPB posits that attitudes (students' positive or negative evaluations), subjective norms (perceived social pressures), and perceived behavioral control (confidence in managing ChatGPT use ethically) directly influence behavioral intention to use ChatGPT (Ahadzadeh et al., 2024; Al-Qaysi et al., 2025; Tran et al., 2024). In this integrated framework, moral disengagement mechanisms are conceptualized as distal predictors that influence the three TPB components, which subsequently predict behavioral intention. Accordingly, six hypotheses were proposed: (H1) moral disengagement mechanisms predict attitudes; (H2) moral disengagement mechanisms predict subjective norms; (H3) moral disengagement mechanisms predict perceived behavioral control; (H4) attitudes predict behavioral intention; (H5) subjective norms predict behavioral intention; and (H6) perceived behavioral control predicts behavioral intention.

## METHODOLOGY

### Participants and Procedure

The sample included 418 college students from different academic programs in the Philippines. All participants had prior experience using ChatGPT for academic writing. Their academic tasks included assignments, feasibility studies, case studies, and theses. Participation was voluntary, and informed consent was obtained. The survey was conducted online using institutional platforms and social media. The study followed a cross-sectional design and collected students' self-reported data on their perceptions, attitudes, and behaviors related to ChatGPT use.

The students had a mean age of 21.65 years. The sample included both male and female respondents. Most of them studied in public higher education institutions. Students reported different frequencies and purposes for using ChatGPT in their academic tasks. Most students reported regular internet access for academic work. The self-rated academic integrity awareness score was 3.92, which indicates a moderately high level of perceived awareness of academic integrity standards.

**Table 1.**
*Demographic profile (N = 418)*

| Variable | n | % |
| --- | --- | --- |
| Sex | | |
| - Male | 245 | 58.5 |
| - Female | 174 | 41.5 |
| Enrolled in | | |
| - Public | 359 | 85.7 |
| - Private | 60 | 14.3 |
| Type of Scholarly Work* | | |
| - Thesis | 143 | 34.1 |
| - Feasibility study | 95 | 22.7 |
| - Case analysis | 100 | 23.9 |
| - Capstone project | 147 | 35.1 |
| - Research report or term paper | 166 | 39.6 |
| - Market study | 64 | 15.3 |
| - Action research | 59 | 14.1 |
| - Others | 35 | 8.37 |
| Frequency of using ChatGPT for academic work | | |
| - Once | 9 | 2.1 |

| Variable | n | % |
|---|---|---|
| -     Rarely | 87 | 20.8 |
| -     Monthly | 101 | 24.1 |
| -     Weekly | 175 | 41.8 |
| -     Daily | 47 | 11.2 |
| Purpose of using ChatGPT* | | |
| -     Ideas | 342 | 81.6 |
| -     Drafts | 154 | 36.8 |
| -     Rewriting/ editing | 192 | 45.8 |
| -     Referencing | 123 | 29.4 |
| -     Summarizing | 191 | 45.6 |
| Regular access to internet for academic work | | |
| -     Yes | 361 | 86.2 |
| -     No | 58 | 13.8 |

*Note*: *Multiple answers

## Instrument

The survey measured nine constructs based on moral disengagement theory and the theory of planned behavior, all contextualized to students' use of ChatGPT for academic writing. All items used a five-point Likert scale where 1 indicated strong disagreement and 5 indicated strong agreement. Moral justification (3 items) measured students' beliefs that using ChatGPT is acceptable if they perceive educational benefits or learning intentions. Euphemistic labeling (3 items) measured how students framed ChatGPT use as academic support rather than misconduct. Displacement of responsibility (3 items) measured how students shifted responsibility to others, such as instructors or peers. Minimizing consequences (3 items) measured beliefs that using ChatGPT causes minimal harm or does not violate academic integrity as long as students understand or modify the content. Attribution of blame (3 items) measured how students assigned responsibility to institutional policies, unclear guidelines, or external pressures. Attitudes toward ChatGPT use (3 items) measured students' positive or negative evaluations of using ChatGPT in academic writing. Subjective norms (3 items) measured students' perceptions of peer influence and social approval regarding ChatGPT use in academic tasks. Perceived behavioral control (3 items) measured students' confidence in their ability to use ChatGPT ethically and manage its use appropriately. Behavioral intention (3 items) measured students' intentions to continue using ChatGPT in future academic writing activities. The instrument also collected demographic information, including degree program, year level, and the type of academic work where ChatGPT was used. The instrument underwent pilot testing to ensure clarity and appropriateness of the items. Reliability testing showed that all constructs achieved acceptable internal consistency, with Cronbach's alpha values greater than .70.

## Data Analysis

Composite scores were calculated by averaging the items under each construct. All variables were standardized before analysis to obtain standardized coefficients. The study used multiple linear regression analyses to examine the hypothesized relationships. Four separate regression models were conducted. The first model tested whether moral disengagement mechanisms predicted attitudes. The second model tested whether moral disengagement mechanisms predicted subjective norms. The third model tested whether moral disengagement mechanisms predicted perceived behavioral control. The fourth model tested whether attitudes, subjective norms, and perceived behavioral control predicted behavioral intention. All regression models used ordinary least squares estimation. The analysis was conducted using Python in Jupyter Notebook.

# RESULTS AND DISCUSSION

## Descriptive Statistics

Table 2 presents the overall mean and standard deviation for each construct. The students generally agreed that using ChatGPT is acceptable if they still include their original thinking (M = 3.78, SD = 1.06), that it is not unethical if ChatGPT helps explain ideas better (M = 3.49, SD = 1.07), and that using ChatGPT is fine if the goal is to learn, not to cheat (M = 3.93, SD = 1.10). They considered ChatGPT a tool for support rather than cheating (M = 3.91, SD = 1.11), viewed asking ChatGPT for ideas like discussing with a classmate (M = 3.73, SD = 1.10), and agreed that rewording ChatGPT's output makes it acceptable to use (M = 3.41, SD = 1.09). The respondents reported that many students in their class use ChatGPT, so they also use it (M = 3.36, SD = 1.15), rely on ChatGPT more when instructors do not provide clear rules (M = 3.22, SD = 1.13), and do not feel fully responsible for using ChatGPT since it is widely used (M = 3.12, SD = 1.04). They stated that using ChatGPT is not a serious issue if they understand the content (M = 3.63, SD = 1.04), that AI-assisted writing is less harmful than copying from another student (M = 3.39, SD = 1.09), and that it is a small issue to use ChatGPT if the work remains mostly theirs (M = 3.54, SD = 1.04).

The participants reported that students use ChatGPT more when schools give unclear policies (M = 3.44, SD = 1.07), that they use ChatGPT more when the academic workload becomes overwhelming (M = 3.59, SD = 1.05), and that the teachers should guide students better if AI use is discouraged (M = 3.70, SD = 1.09). They agreed that ChatGPT helps improve the quality of academic work (M = 3.61, SD = 1.03), that it makes writing tasks easier (M = 3.54, SD = 1.05), and that they feel positive about using ChatGPT in school assignments (M = 3.37, SD = 1.10). Many participants agreed that students they know use ChatGPT for academic writing (M = 3.84, SD = 1.05), that using ChatGPT for projects is common in their school (M = 3.69, SD = 1.09), and that they feel social pressure to use ChatGPT because others do (M = 3.09, SD = 1.15). They reported that they know how to use ChatGPT without violating school rules (M = 3.70, SD = 1.06), feel confident in using ChatGPT ethically (M = 3.52, SD = 1.04), and can recognize when ChatGPT use becomes academically dishonest (M = 3.62, SD = 1.02). The respondents expressed intentions to use ChatGPT in future academic activities (M = 3.35, SD = 1.07), to continue using ChatGPT even for major requirements (M = 3.25, SD = 1.13), and to use ChatGPT unless strict rules prohibit it (M = 3.51, SD = 1.11).

**Table 2.**
*Descriptive statistics of constructs*

| Construct | Mean | SD |
|---|---|---|
| Moral Justification | 3.73 | 0.91 |
| Euphemistic Labeling | 3.68 | 0.95 |
| Displacement of Responsibility | 3.24 | 0.92 |
| Minimizing Consequences | 3.52 | 0.94 |
| Attribution of Blame | 3.58 | 0.91 |
| Attitudes Toward ChatGPT Use | 3.51 | 0.99 |
| Subjective Norms | 3.54 | 0.9 |
| Perceived Behavioral Control | 3.62 | 0.94 |
| Behavioral Intentions | 3.37 | 1.03 |

## Predictors of Attitudes Toward ChatGPT Use

A multiple regression analysis examined the extent to which moral justification, euphemistic labeling, displacement of responsibility, minimizing consequences, and attribution of blame predicted students' attitudes toward ChatGPT use. The model was statistically significant, $F(5, 412) = 173.90$, $p < .001$, accounting for approximately 67.9% of the variance ($R^2 = .679$). Euphemistic labeling (β = 0.133, p = .010), displacement of responsibility (β = 0.117, p = .002), minimizing consequences (β = 0.313, p < .001), and attribution of blame (β = 0.319, p < .001) were significant predictors, while moral justification was not significant (β = 0.068, p = .178).

These results suggest that students who soften the ethical implications of using ChatGPT, transfer responsibility to others, downplay possible academic harm, or assign fault to external academic structures tend to hold more favorable attitudes toward using ChatGPT for academic writing (Acosta-Enriquez, Arbulú Ballesteros, Arbulu Perez Vargas, et al., 2024; Acosta-Enriquez, Arbulú Ballesteros, Huamaní Jordan, et al., 2024). Minimizing consequences and attribution of blame produced the strongest effects. Students likely felt more comfortable using ChatGPT when they believed that institutional guidelines were unclear or the consequences of use were minor (Bikanga Ada, 2024; Hasanein & Sobaih, 2023). Moral justification did not predict attitudes, suggesting that students may not rely heavily on internal justifications but instead use external rationalizations.

### Predictors of Subjective Norms

The regression model predicting subjective norms was significant, $F(5, 412) = 107.80$, $p < .001$, explaining 56.7% of the variance ($R^2 = .567$). Euphemistic labeling (β = 0.125, p = .038), displacement of responsibility (β = 0.149, p = .001), minimizing consequences (β = 0.169, p = .002), and attribution of blame (β = 0.387, p < .001) were significant predictors. Moral justification was not significant (β = 0.038, p = .521).

The findings indicate that students' perceptions of social acceptance for ChatGPT use relate closely to their moral disengagement rationalizations (Zhang et al., 2025). Attribution of blame had the strongest effect. Students who believed that institutions or instructors failed to provide clear guidance perceived stronger peer acceptance for ChatGPT use (Santos et al., 2024; Strzelecki, 2024). Euphemistic labeling, displacement of responsibility, and minimizing consequences also influenced subjective norms, suggesting that external rationalizations contributed to students' sense that ChatGPT use was common and socially supported. Again, moral justification did not predict subjective norms, which shows that peer norms depend more on external explanations rather than internal ethical framing.

### Predictors of Perceived Behavioral Control

The predictors also explained significant variance in perceived behavioral control, $F(5, 412) = 119.00$, $p < .001$, with $R^2 = .591$. Moral justification (β = 0.258, p < .001), minimizing consequences (β = 0.124, p = .017), and attribution of blame (β = 0.421, p < .001) emerged as significant predictors. Euphemistic labeling and displacement of responsibility were not significant.

These results show that students who assign responsibility to institutions, minimize potential harm, or frame ChatGPT use as beneficial feel more confident in their ability to manage AI-assisted writing ethically (Blahopoulou & Ortiz-Bonnin, 2025; Cheng et al., 2025; Grimes & Barton, 2024). Attribution of blame served as the strongest predictor. This also means that students who blamed academic structures for lack of clarity or oversight felt more capable of using ChatGPT without crossing ethical boundaries. Minimizing consequences and moral justification also raised students' perceived control by reducing the perceived risks or ethical concerns associated with ChatGPT use (Howlader et al., 2025; Zhang et al., 2025).

### Predictors of Behavioral Intention

The attitudes, subjective norms, and perceived behavioral control significantly predicted behavioral intention to use ChatGPT, $F(3, 414) = 212.10$, $p < .001$, explaining 60.6% of the variance ($R^2 = .606$). Attitudes (β = 0.499, p < .001), subjective norms (β = 0.098, p = .037), and perceived behavioral control (β = 0.253, p < .001) all significantly predicted behavioral intention (Fig. 1).

The results confirm that students' behavioral intentions to use ChatGPT depend most strongly on their favorable evaluations of ChatGPT use. Attitudes served as the strongest predictor, which supports the Theory of Planned Behavior's assumption that attitudes are the most immediate determinant of behavioral intention. Subjective norms and perceived behavioral control also influenced intention. Students who perceived peer support and who felt confident in managing ChatGPT use expressed stronger intentions to continue using the tool for academic writing.

**Figure 1.**
*Final structural model with standardized estimates*

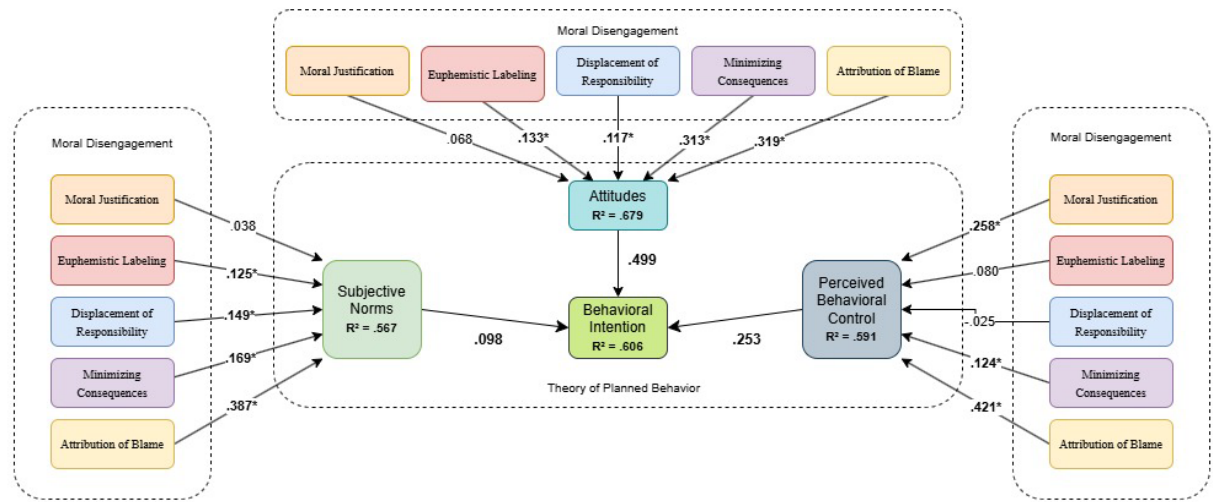

# CONCLUSION

This study examined how moral disengagement influences the intention of Filipino college students to use ChatGPT in academic writing. The results showed that students who blamed institutions, minimized possible consequences, or shifted responsibility were more likely to form positive attitudes and feel confident using ChatGPT. Attitudes had the strongest influence on behavioral intention, followed by perceived control and social norms. These results suggest that students' decisions to use AI tools are shaped by both their moral reasoning and the academic environment around them.

These results raise ethical concerns that schools must address. Many students think it is acceptable to use ChatGPT if the purpose is learning or if they revise the content. Some students justify their actions by pointing to unclear school policies. This situation shows that many students do not have a clear understanding of academic honesty. Schools should update and clearly define academic honesty, especially in the context of AI use. They should not only provide rules but also give proper guidance to help students reflect on the moral aspects of using AI tools. Teachers can support this effort by leading classroom discussions, using real examples, and giving feedback. A strong academic culture that promotes honesty, accountability, and responsibility can help students make better decisions when using AI in their academic work.

This study also shows that existing models like the TPB do not fully explain student behavior in real situations. The model assumes that students always act based on logic, but other factors like stress, pressure, or ease of access can lead to choices that are not fully planned. Emotional and cultural factors also play a role in how students view fairness and responsibility. Future studies should explore these areas to build better models that match how students behave in digital learning environments. This study supports the need for clear rules, faculty guidance, and programs that develop moral responsibility in students who use AI tools.